\documentclass[aps,prd,twocolumn]{revtex4}
\usepackage{graphicx}

\begin{document}

\title{Fissile isotopes antineutrino spectra: summation method and direct experiment}

\author{P.~Naumov$^{2}$, S.~Silaeva$^{1}$, V.~Sinev$^{1,2}$, A.~Vlasenko$^{1,2}$}
%\email{vsinev@inr.ru}

\affiliation{$^{1}$ Institute for Nuclear Research of Russian Academy of Sciences (INR RAS), Moscow, Russia}
\affiliation{$^{2}$ National Research Nuclear University MEPhI (Moscow Engineering Physics Institute), Moscow, Russia}

\begin{abstract}
New antineutrino spectra of fissile isotopes ($^{235}$U, $^{238}$U, $^{239}$Pu and $^{241}$Pu) which are containing in a nuclear reactor fuel have been obtained. A combined technique was used: calculation of antineutrino spectra and their fitting to those obtained in the experiment at the Rovno NPP in the 80s of the last century. The cross sections of fissile isotopes calculated with these spectra describe well the cross section obtained experimentally in the Double Chooz experiment. The calculated cross section for the same reactor core composition is $\sigma_{f} = (5.82 \pm 0.12)\times10^{-43}$ cm$^2$/fission. It is the closest result between all predicted. The obtained spectra have the same "bump" as experimental ones in 5 MeV observed energy region.
\end{abstract}
\maketitle

\section{Introduction}
The history of the antineutrino spectrum of a nuclear reactor goes back more than half a century. Antineutrinos from nuclear reactor are registered with highest efficiency using the reaction of inverse beta-decay (IBD)
\begin{equation}
\bar{\nu}_e + p \rightarrow n + e^+.
\end{equation}
The positron in this reaction takes all the energy of the antineutrino minus the reaction threshold of 1.806 MeV.

While the statistics in reactor experiments amounted to tens of thousands of neutrino events, the calculated spectrum satisfactorily described the experimentally observed spectrum of IBD positrons \cite{zacek}$–$\cite{reines}. However, in recent experiments on the search for the neutrino mixing angle $\theta_{13}$, the statistics already amounted to millions of events, and the discrepancy between the measured and calculated spectra was clearly manifested. A peak appeared in the region of 6 MeV (5 MeV in the observed energy) in the antineutrino energy, which cannot be obtained by calculation \cite{kerret}$–$\cite{an}. The cross section for the IBD reaction using the calculated spectrum \cite{muell}, \cite{huber} turns out to be approximately 3\% larger than the experimental one, which is now associated with the presence of sterile neutrinos.

In addition to the calculated spectra for three isotopes ($^{235}$U, $^{238}$U and $^{239}$Pu) fissile by thermal neutrons, antineutrino spectra were obtained by measuring the beta spectra of fission fragments in the ILL experiments in 1982-1989 \cite{schre}, \cite{hahn}. In these spectra, there is a slight rise in the region of 6 MeV, which falls short of that observed in experiments. This may be due to some incorrectness in the measured beta electrons spectrum conversion into antineutrinos.

In 1990, the first experimental spectrum of antineutrinos was obtained in an experiment at the Rovno NPP \cite{klimov}. This spectrum corresponded to a certain composition of a nuclear reactor core and was obtained in the form of a formula describing the spectrum on average. Later, a technique was developed that made it possible to convert the measured IBD reaction positron spectrum into an antineutrino one, which could be splitted into constituent components $-$ the spectra of four fissile isotopes ($^{235}$U, $^{238}$U, $^{239}$Pu and $^{239}$Pu) in \cite{sinev}.

It would seem that the spectrum of antineutrinos from a fissile isotope is easy to calculate, knowing the probabilities of the birth of fragments, but in practice everything turns out to be more complicated. Short-lived fragments are mostly located far from the beta stability line. Many of them do not know how to decay, and their number reaches a quarter of all fragments. So, the calculated spectrum of antineutrinos does not describe the spectrum observed in the experiment with large statistics.

In the first calculations of antineutrino spectrum \cite{kopeik}$–$\cite{rubts} from uranium fission, about 500 fragments were used, distributed along a double-humped fragment mass distribution curve. The fragment yields were described using the Gaussian function inside the charge chain for a given fragment mass. Currently, the databases of fission fragments contain more than a thousand nuclei with masses from 58 to 184, and taking into account ternary fission, light nuclei with masses from 1 to 15 are also added. However, data for fragments on direct fission yield and decay schemes remain inaccurate and in present time. It is extremely difficult to determine the fission yield of a fragment if its lifetime is much less than one second, not to mention the probabilities of beta transitions of the fragment to the daughter nucleus. Basically, the data on beta transitions of short-lived nuclei are estimated by analogy with the known long-lived ones. Out of 948 nuclei (together with excited states) used in our calculation, 385 nuclei can be considered known, including 92 stable ones. Another 231 are assessed and 332 are unknown.

In this paper, we present a new calculation of the antineutrino spectra of fissile isotopes based on our upgraded database of fission fragments. In the available database for fragments with unknown decay schemes, the strength function was used to describe the probabilities of nuclear beta transitions. The strength function was chosen to describe better the antineutrino spectra obtained in the Rovno experiment \cite{sinev}.

\section{Summation method of antineutrino spectrum calculation}

In a nuclear reactor core, $\bar{\nu_e}$-s are produced by $\beta$-decays of neutron-exceeded fission fragments. Fragments, on the other hand, appear as a result of the fission reaction of heavy nuclei of uranium and plutonium, which make up the fuel of a nuclear reactor. In this case, the set of fragments is characteristic for each fissile isotope. Accordingly, the antineutrino spectra will be individual for each nucleus undergoing fission.

In the introduction, it was said that the antineutrino spectrum of a nuclear reactor is formed by four isotopes of uranium and plutonium. In the calculation, the spectra of these four isotopes are obtained independently. The reactor $\bar{\nu_e}$-s spectrum at any moment of time is a superposition of the spectra of these isotopes with weights corresponding to their fission fractions in the core of nuclear reactor.

\begin{figure}[ht]
\begin{center}
\includegraphics[width=86mm]{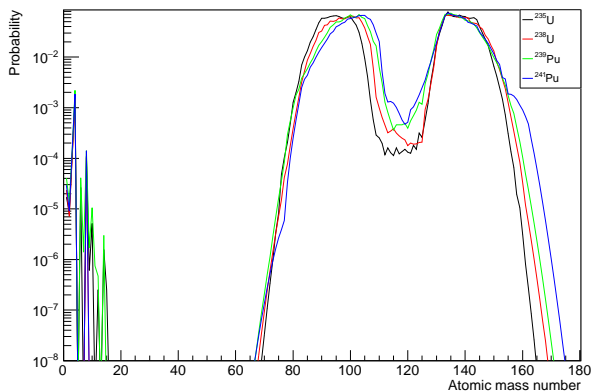}
\end{center}
\caption{\label{fig:fig1}Distribution of fragment yields per fission event for four fissile isotopes: $^{235}$U $-$ black line, $^{238}$U $-$ red line, $^{239}$Pu $-$ green line and $^{241}$Pu $-$ blue line, which are in the composition of nuclear fuel. The data are given according to the database \cite{iaea}. In the range of small fragment masses, light nuclei are observed that take part in triple fissions.}
\end{figure}

The calculation method consists in summing the individual spectra $\bar{\nu_e}$ from all fission products, taking into account their yields in the fission process. The spectrum of antineutrinos or beta particles from fission products of nuclear fuel in a reactor is described by the expression:

\begin{equation}
f_{\nu}(E) = \sum_{j,k} Y_{j}\cdot b_{j,k}\cdot S_{j,k}(E),
\end{equation}
where $S_{j,k}(E) -$ is individual fragment antineutrino spectrum, $b_{j,k} -$ is probability of beta-transition and $Y_{j} -$ fragment direct yield that is probability to be born directly in a fission. Summation is made through all fragments.

The individual beta spectrum shape $P_{e}(E_{e},E_{0},Z)$ can be written like this:

\begin{eqnarray}
P_{e}(E_{e},E_{0},Z) = K\cdot p_{e}E_{e}\cdot (E_{0}-E_{e})^2\cdot F(Z,E_{e})\cdot \nonumber \\
C(Z,E_{e})\cdot(1+\delta(Z,A,E_{e})),
\end{eqnarray}
where $K -$ is normalization factor, $p_{e}$ and $E_{e} -$ are momentum and energy of electron, $F(Z,E_{e}) -$ Fermi function accounting Coulomb field of the daughter nucleus, $C(Z,E_{e}) -$ the factor accounting momentum dependence of nucleus matrix element and $\delta(Z,A,E_{e}) -$ is correction factor to the spectrum shape.

The antineutrino spectrum shape $P_{\bar{\nu}}(E_{\bar{\nu}},E_{0},Z)$ can be expressed the same way as beta spectrum one by changing $E_e$ with $E_0 - E_{\bar{\nu}}$.

To calculate the antineutrino spectrum, we used the IAEA database \cite{iaea} on fission fragments. This database is compiled from several nuclear databases and can be considered as most complete. There are about 1050 fragments involved in the creation of the antineutrino spectrum in total. Of these, 332 have unknown decay schemes and an estimated half-life, which is much less than one second. Basically, these fragments have a low fission yield, but a high beta transition energy ($Q_{\beta}$).

The database contains data on the probability of the birth of a nucleus with mass A and charge Z in the fission of a number of heavy nuclei. The probability of the birth of a nucleus is called the direct yield of a fragment during fission. If we sum up the direct yields of fragments preceding the one selected in the beta decay chain, then we get the cumulative yield of this fragment. The cumulative yields for stable isotopes of all fragments are shown in Figure~\ref{fig:fig1}.
The database included light fragments ($A$ = 1 to $A$ = 15) from hydrogen to carbon, which contribute to the antineutrino spectrum in the case of ternary fission.

\section{Experimental antineutrino spectrum that follows from Rovno experiment}

In the late 1980s, an experiment was carried out at the Rovno NPP to measure the antineutrino spectrum using a small (by today's standards) detector (sensitive volume $\sim$0.5 m$^3$) \cite{klimov}. For three years of measurements (from 1988 to 1990), 174 thousand neutrino events was collected. The IBD reaction (1) was used to detect antineutrinos.

The positron kinetic energy T of reaction (1) in the first approximation is approximately equal to the antineutrino energy E minus the threshold energy $\Delta$ and the neutron recoil energy $T\approx E – \Delta – r_{n}$ ($\Delta$ = 1.806 MeV, and $r_{n}$ is less than 20 keV).
The experimental positron spectrum is a convolution of the antineutrino spectrum with the inverse beta decay reaction cross section and the detector response function

\begin{equation}
S_{e}(T) = \int \rho_{\nu}(E)\cdot \sigma_{\nu,p}\cdot (E)R(T,E)dE,
\end{equation}
where $S_{e}(T) -$ positron spectrum, $\rho_{\nu}(E) -$ antineutrino spectrum,$R(T,E) -$ detector response function.

To solve equation (4) and get the antineutrino spectrum, the calculated detector response function obtained by modeling the process of positron registration in the detector for a number of positron energies was used. The antineutrino spectrum looked for as an exponential function with a 10th degree polynomial in the exponent. Three terms were left in the polynomial: the first, second and tenth degrees. Thus, four parameters were used to describe the antineutrino spectrum: a normalization factor and three coefficients at the antineutrino energy degrees:

\begin{equation}
\rho_{\nu}(E) = C_{1}\cdot exp(C_{2}E + C_{3}E^{2} + C_{4}(E/8)^{10}).
\end{equation}
The coefficients of the function describing the behavior of the antineutrino spectrum (5) were found: $C_{1}$ = 5.09, $C_{2}$ = –0.648, $C_{3}$ = –0.0273, $C_{4}$ = –1.411. This spectrum corresponds to a certain composition of the nuclear reactor core in terms of fission fractions of heavy isotopes: $\alpha$($^{235}$U) = 0.586, $\alpha$($^{238}$U) = 0.075, $\alpha$($^{239}$Pu) = 0.292, $\alpha$($^{241}$Pu) = 0.047.
The positron spectrum conversion into an antineutrino one procedure was developed later in \cite{sinev}. In this procedure, the spectrum observed in the detector was transformed from observed energy into the antineutrino one, then the influence of the detector was excluded from it, and it was divided by the IBD reaction cross section. Then, the antineutrino spectra of individual isotopes of nuclear fuel ($^{235}$U, $^{238}$U, $^{239}$Pu and $^{241}$Pu) were isolated from the reactor spectrum, according to a known core fuel composition.

The $^{235}$U antineutrino spectrum turned out to be similar in shape to the calculated spectra of other authors \cite{schre}, \cite{hahn}. The difference was in presence of a bulge in the region of 6 MeV, which corresponds to the observed anomaly in the antineutrino spectra at 5 MeV observed energy in all reactor experiments. Also, an increased part of the spectrum was observed in the region below 2.5-3 MeV down to the IBD reaction threshold (1.806 MeV), what may be due to the detection of antineutrinos from the spent fuel pool located directly near the nuclear reactor and incomplete removal energy resolution of the detector.

The antineutrino spectra obtained in \cite{sinev} can be called experimental, since the method for converting the observed IBD reaction positron spectrum into an antineutrino one was implemented for the first time.

\section{Experimental antineutrino spectra fit by calculation}

As was shown above, the number of fission fragments with unknown decay schemes is quite large and amounts to approximately one third of all fragments. This is expected, because it is difficult to obtain experimentally beta spectra of nuclei with a decay time of less than one second. In many cases, the decay patterns and beta transition probabilities of such nuclei are estimated using an analogy with similar nuclei that have a longer half-life or with nuclei that can be obtained by irradiating stable isotopes with neutrons and having similar parity and moment.

We decided to use nuclei with unknown decay schemes in order to achieve better agreement between the calculated antineutrino spectra and ones from data obtained in experiments at Rovno \cite{klimov} in \cite{sinev}.

The calculation method described above uses a database that includes fragments with unknown decay patterns. They usually have a large beta decay energy $Q_{\beta}$, but the probabilities of beta transitions are unknown. When calculating, a simplified decay scheme is usually used: with one, or two or three levels of the daughter nucleus. We have replaced this approach with the use of a multi-level system of the daughter nucleus, a kind of strength function. The probabilities of beta transitions were distributed according to the Gaussian function with an average value lying in the energy range from 2 MeV to $Q_{\beta}$ and a dispersion equals to 20\% of the mean energy. The mean energy was selected as a parameter equal to some fraction of $Q_{\beta}$. The selection of the average energy was carried out by minimizing the functional composed of the calculated and experimental spectra.

\begin{equation}
\chi^{2} = \sum_{k=5,8,9,1}\sum_{i=1}^{23} \Big(\frac{^{exp}y_{j,k}-^{calc}y_{j,k}}{\sigma_{j,k}}\Big)^2,
\end{equation}
where $^{exp}y_{j,k} -$ experimental antineutrino spectrum from \cite{sinev}, $^{calc}y_{j,k} -$ calculated spectrum and $\sigma_{j,k} -$ uncertainty of experimental spectrum.

A part of the experimental spectra above 3.5 MeV was selected for fitting to cut off the low-energy part, where the spectrum from spent nuclear fuel located in the settling pool next to the nuclear reactor could be present, and which could increase the experimental spectrum compared to the pure spectrum from fission fragments. At each stage of minimization, a new calculated spectrum was created with a change in the mean energy of beta transitions of unknown fragments.

When calculating antineutrino spectra, the question arises: how to calculate correctly the spectrum? Is it necessary to take into account the Fermi function in formula (3) for the antineutrino spectrum? The neutrino is a neutral, ultrarelativistic particle, and it should not interact with the electric fields of the nucleus and electron shells, unlike the electron. According to modern concepts, beta decay occurs through the emission of a $W$-boson by a $d$-quark, which in turn decays into an electron and an antineutrino. If the $W$-boson had time to fly out of the atom before decay, then it would be necessary to take into account the Fermi function for antineutrinos, but if not, then it is not necessary. The lifetime of the $W$-boson is about $10^{–25}$ s, during which time the $W$-boson has time to cover a path of less than one nucleon radius. If the Fermi function is not used to calculate the antineutrino spectrum, the individual spectra of the antineutrino and electron turn out to be not exactly mirrored. The question of the need to check the symmetry of the beta and antineutrino spectra was raised in \cite{silaeva}.

We have done minimization for two cases: using the Fermi function, and without using the Fermi function. It was shown above that the fragment database contains almost in equal proportions the known, unknown, and estimated data on beta transitions. The estimated data can be correct with some degree of probability. The strength function was applied to both unknown and estimated fragments, assuming that they were also unknown.

We have obtained the following values of $\chi^{2}$ corresponding to cases described above.
For all four antineutrino spectra, 92 bins were used, 23 bins per spectrum. The best agreement between the calculated and experimental spectra is observed when the Fermi function is not used and the strength function is applied to both unknown and estimated fragments. This indirectly says in favor of the non-symmetry of beta and antineutrino spectra. An attempt to describe the experimental spectra by varying the decay schemes of only unknown fragments and using the Fermi function led to a value of ${\chi}^{2}$ many times greater. The result of the fitting is in the Table \ref{tabl:chi2}.

\begin{table}[ht]
\caption{The result of strength function application for the unknown and estimated fragments with and without Fermi function in the expression for antineutrino spectrum shape (3).}
\label{tabl:chi2}
\centering
\vspace{2mm}
\begin{tabular}{ c | c | c }
\hline
\hline
 & $F(Z,A)$ & $\chi^{2}$ \\
\hline
Unknown & "+" & 2598 \\
 & "$-$" & 892 \\
\hline
Unknown and & "+" & 290 \\
estimated & "$-$" & 95 \\

\hline
\hline
\end{tabular}
\end{table}

\begin{figure}[ht]
\centering
\begin{minipage}{0.44\linewidth}
\includegraphics[width=\linewidth]{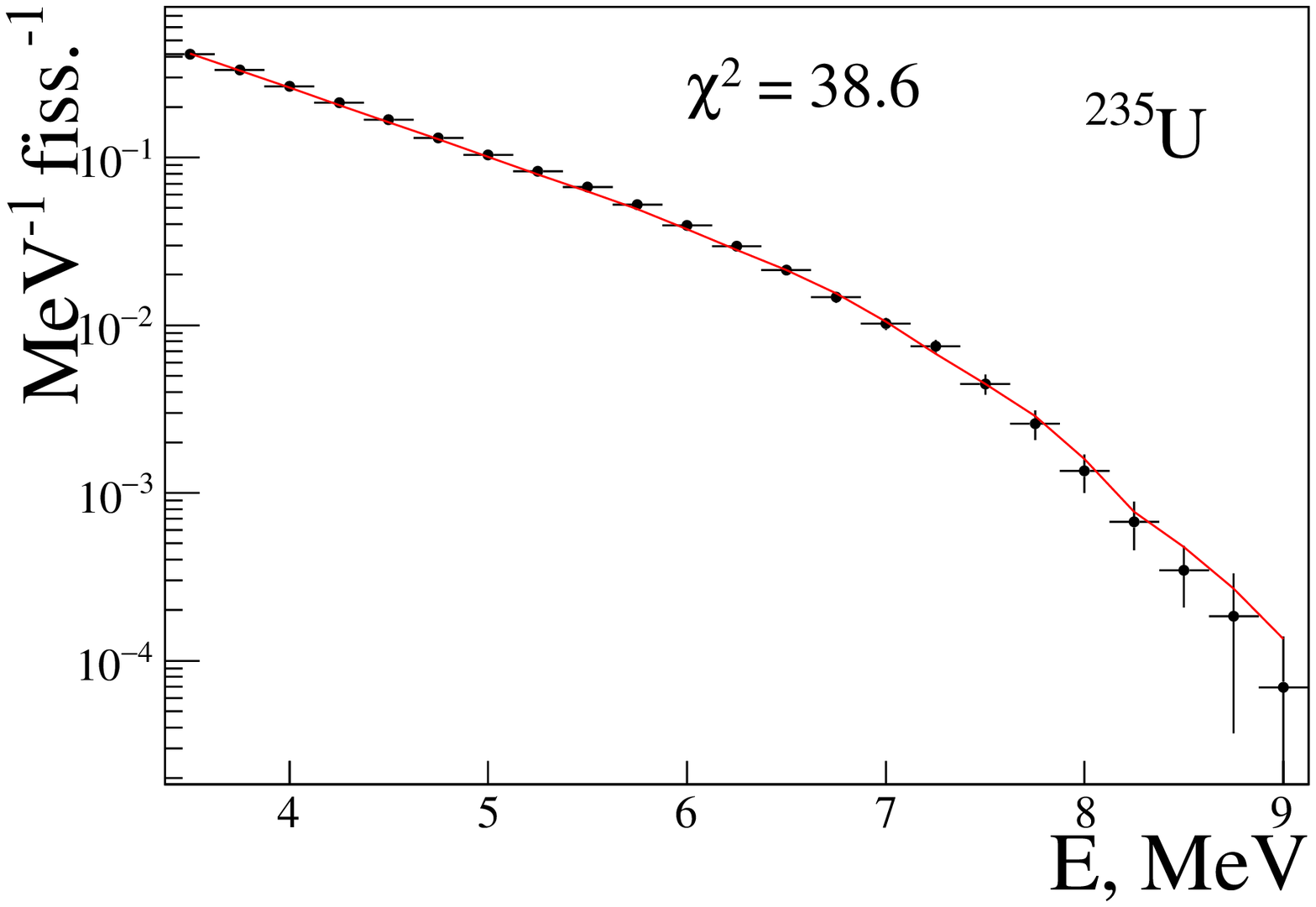}
\end{minipage}
\begin{minipage}{0.44\linewidth}
\includegraphics[width=\linewidth]{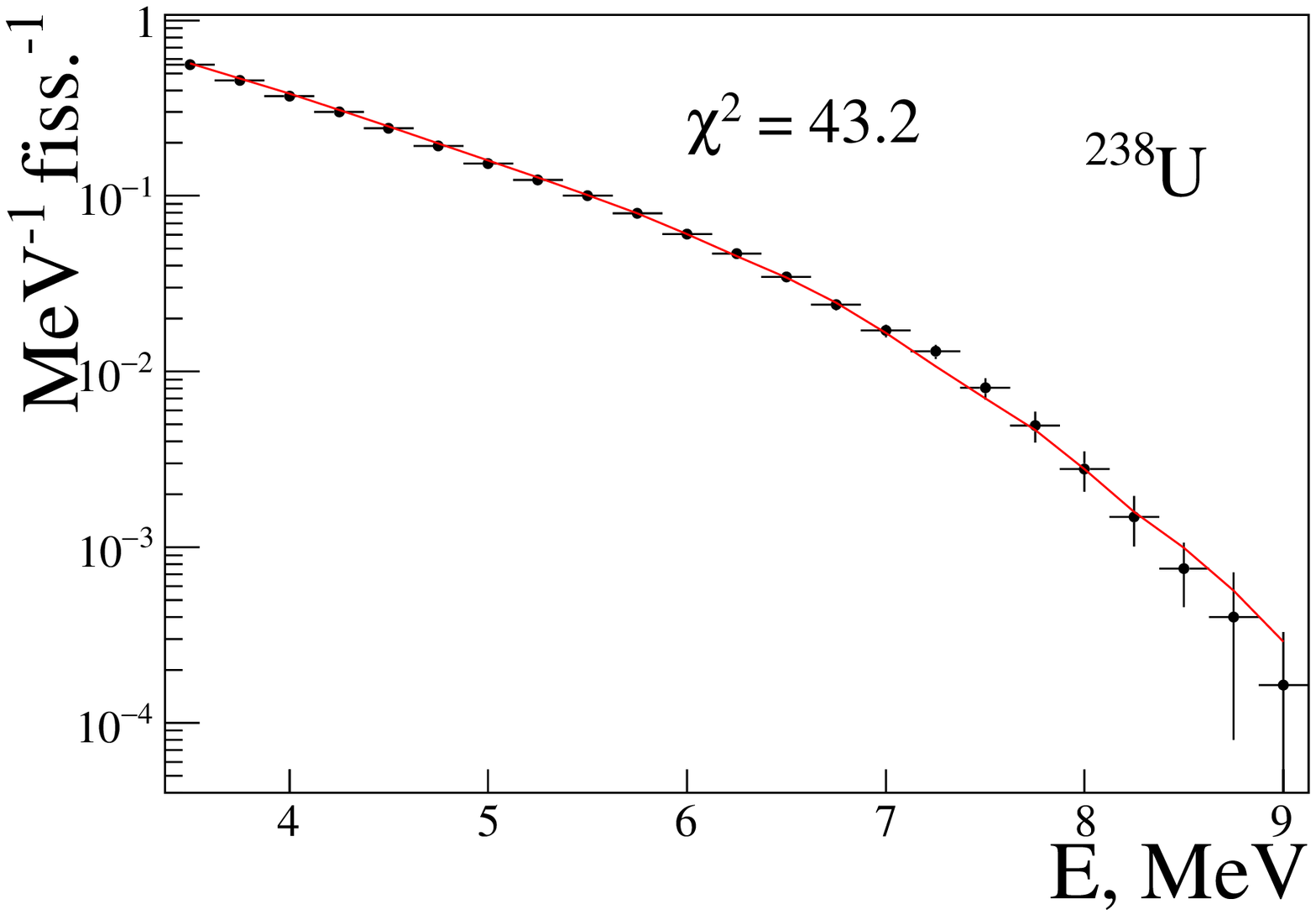}
\end{minipage}
\begin{minipage}{0.44\linewidth}
\includegraphics[width=\linewidth]{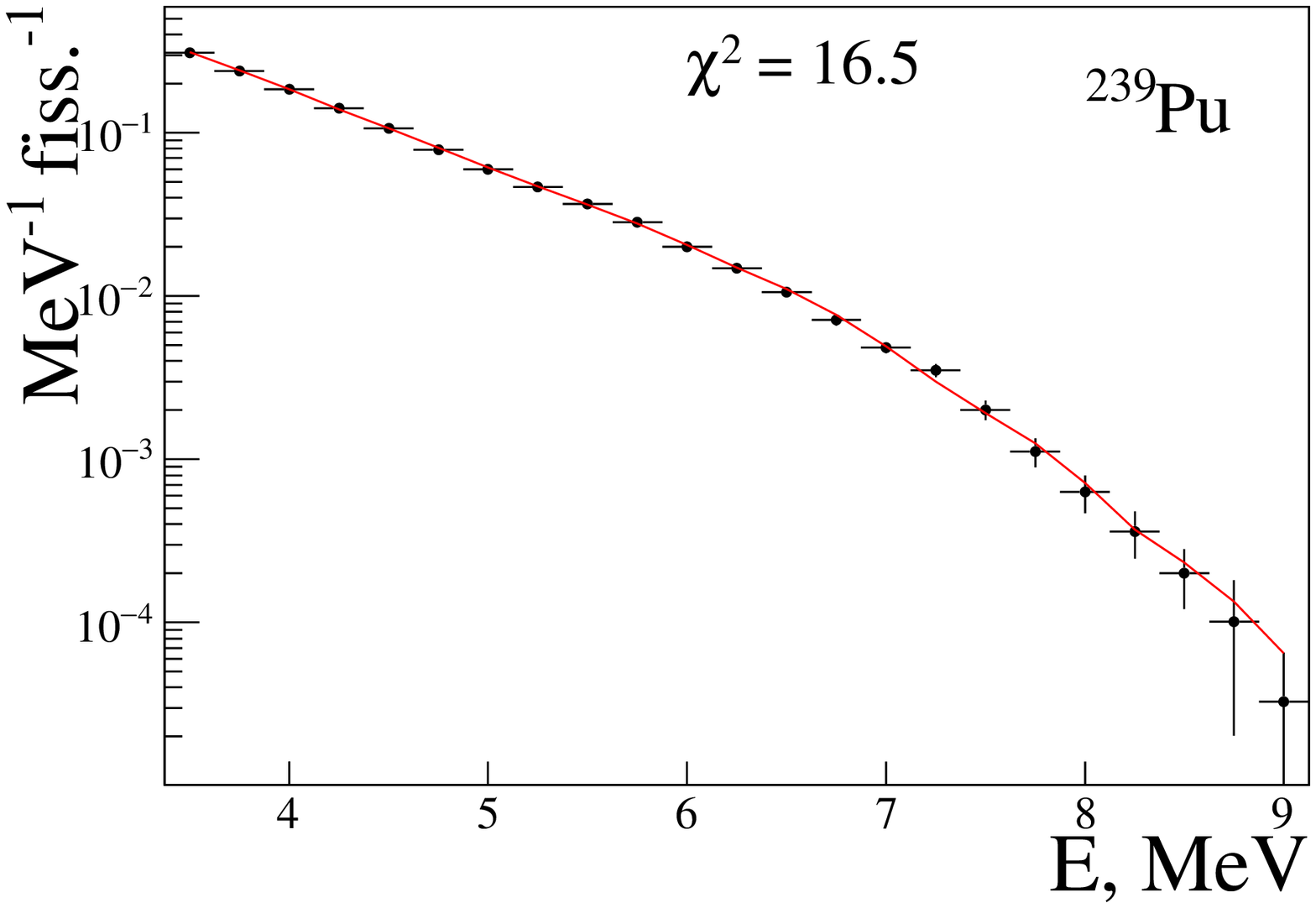}
\end{minipage}
\begin{minipage}{0.44\linewidth}
\includegraphics[width=\linewidth]{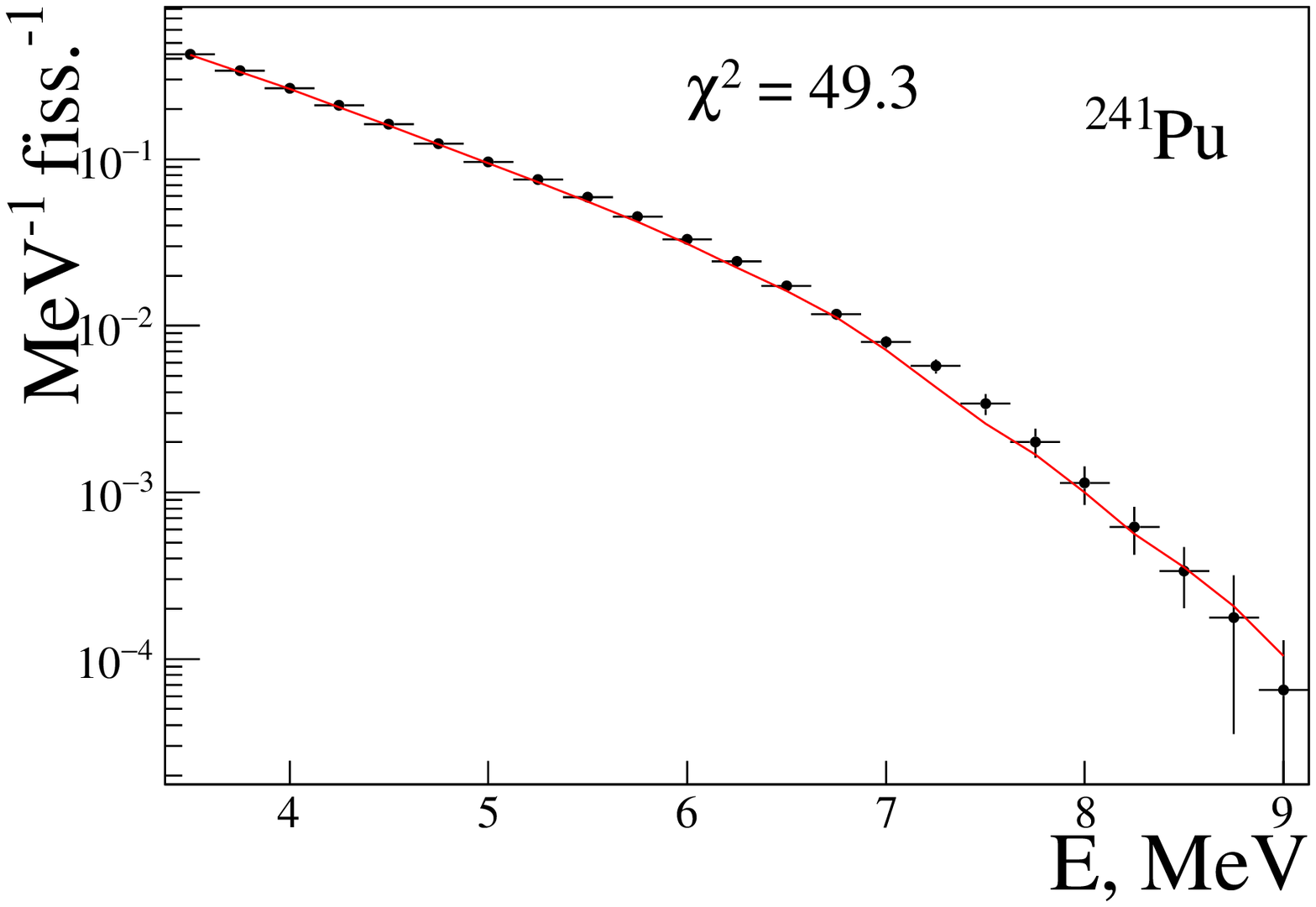}
\end{minipage}
%\end{center}
\caption{\label{figtwo} Fit of experimental antineutrino spectra from \cite{sinev} by calculated ones with using strenght function.}
\end{figure}

As a result of minimization, an upgraded database of fission fragments was obtained. In this base, the probabilities of beta transitions for unknown and estimated fragments are described by a strength function. The new calculated antineutrino spectra obtained by this technique in comparison with experimental spectra \cite{sinev} are shown in Figure~\ref{figtwo}. At start of minimization the discrepancy in calculated was sufficient.

Using the modified base of fission fragments, the antineutrino spectra of all fissile isotopes of nuclear fuel were calculated. These spectra correspond to two years of fuel irradiation in a reactor core. The result is shown in Table \ref{tabl:spectra}.

\begin{table*}[ht]
\caption{Antineutrino spectra of $^{235}$U, $^{238}$U, $^{239}$Pu and $^{241}$Pu for two years of neutron irradiation in reactor core. In brackets the power of 10 is shown for the values.}
\begin{minipage}{\textwidth}
\label{tabl:spectra}
\centering
\vspace{2mm}
\begin{tabular}{ c | c | c | c | c }
\hline
\hline
 $E_{\nu}$, Mev & $^{235}$U & $^{238}$U & $^{239}$Pu & $^{241}$Pu \\
\hline
1.50 & 1.73 & 2.11 & 1.53 & 1.79 \\
1.75 & 1.50 & 1.87 & 1.30 & 1.56 \\
2.00 & 1.29 & 1.61 & 1.08 & 1.32 \\
2.25 & 1.08 & 1.37 & 8.96(-1) & 1.12 \\
2.50 & 8.97(-1) & 1.13 & 7.36(-1) & 9.22(-1) \\
2.75 & 7.51(-1) & 9.55(-1) & 6.08(-1) & 7.64(-1) \\
3.00 & 6.23(-1) & 8.09(-1) & 4.94(-1) & 6.32(-1) \\
3.25 & 5.15(-1) & 6.86(-1) & 4.00(-1) & 5.22(-1) \\
3.50 & 4.16(-1) & 5.70(-1) & 3.14(-1) & 4.21(-1) \\
3.75 & 3.31(-1) & 4.67(-1) & 2.41(-1) & 3.34(-1) \\
4.00 & 2.62(-1) & 3.81(-1) & 1.85(-1) & 2.64(-1) \\
4.25 & 2.06(-1) & 3.08(-1) & 1.40(-1) & 2.05(-1) \\
4.50 & 1.63(-1) & 2.49(-1) & 1.07(-1) & 1.60(-1) \\
4.75 & 1.29(-1) & 2.00(-1) & 8.11(-2) & 1.23(-1) \\
5.00 & 1.01(-1) & 1.59(-1) & 6.16(-2) & 9.44(-2) \\
5.25 & 7.97(-2) & 1.27(-1) & 4.73(-2) & 7.27(-2) \\
5.50 & 6.30(-2) & 1.01(-1) & 3.65(-2) & 5.59(-2) \\
5.75 & 4.93(-2) & 7.93(-2) & 2.79(-2) & 4.23(-2) \\
6.00 & 3.75(-2) & 6.06(-2) & 2.07(-2) & 3.10(-2) \\
6.25 & 2.82(-2) & 4.55(-2) & 1.50(-2) & 2.23(-2) \\
6.50 & 2.14(-2) & 3.42(-2) & 1.10(-2) & 1.62(-2) \\
6.75 & 1.55(-2) & 2.45(-2) & 7.69(-3) & 1.12(-2) \\
7.00 & 1.05(-2) & 1.65(-2) & 4.91(-3) & 7.12(-3) \\
7.25 & 6.77(-3) & 1.07(-2) & 2.99(-3) & 4.26(-3) \\
7.50 & 4.47(-3) & 7.00(-3) & 1.92(-3) & 2.59(-3) \\
7.75 & 2.87(-3) & 4.64(-3) & 1.26(-3) & 1.69(-3) \\
8.00 & 1.59(-3) & 2.79(-3) & 7.18(-4) & 1.00(-3) \\
8.25 & 7.74(-4) & 1.59(-3) & 3.71(-4) & 5.64(-4) \\
8.50 & 4.75(-4) & 9.87(-4) & 2.34(-4) & 3.56(-4) \\
8.75 & 2.68(-4) & 5.66(-4) & 1.34(-4) & 2.07(-4) \\
9.00 & 1.35(-4) & 2.91(-4) & 6.47(-5) & 1.04(-4) \\
\hline
\hline
\end{tabular}
\end{minipage} \hfill
\end{table*}

%Table \ref{tabl:bor1}

The number of antineutrinos per fission of fissile isotopes (the integral of the antineutrino spectrum) and the IBD reaction cross section for each spectrum (convolution of the antineutrino spectrum with the IBD reaction cross section) are given in Tables \ref{tabl:meannu} and \ref{tabl:sections}. To calculate the IBD reaction cross sections, we used the cross section for a monoenergetic antineutrino taken from \cite{strumia}. Table \ref{tabl:sections} also lists the cross sections for the calculated spectra of a number of other authors.

\begin{table}[ht]
\caption{Mean antineutrino number per fission, $^{fiss}n_{\nu}$}
\label{tabl:meannu}
\centering
\vspace{2mm}
\begin{tabular}{ c | c | c | c }
\hline
\hline
 $^{235}$U & $^{238}$U & $^{239}$Pu & $^{241}$Pu \\
\hline
 5.962 & 7.106 & 5.419 & 6.221 \\
\hline
\hline
\end{tabular}
\end{table}

\begin{table}[ht]
\caption{IBD cross sections calculated up to 9 MeV in antineutrino energy $\sigma_{f}\times10^{-43}$, cm$^{2}$/fission.}
\label{tabl:sections}
\centering
\vspace{2mm}
\begin{tabular}{ c | c | c | c | c | c }
\hline
\hline
 & $^{235}$U & $^{238}$U & $^{239}$Pu & $^{241}$Pu & DC \\
\hline
This work & 6.241 & 9.089 & 4.269 & 5.948 & 5.815 \\
ILL \cite{schre}, \cite{hahn} & 6.395 & 8.903 & 4.185 & 5.768 & 5.840 \\
Vogel \cite{vogel} & 6.498 & 9.135 & 4.508 & 6.520 & 6.066 \\
MEPhI \cite{rubts} & 6.404 & 9.267 & 4.383 & 6.489 & 5.985 \\
Huber-Mueller \cite{muell}, \cite{huber} & 6.658 & 10.08 & 4.364 & 6.031 & 6.154 \\
Kopeikin \cite{kopeik} & 6.308 & 9.395 & 4.33$^*$ & 6.01$^*$ & 5.900 \\
\hline
\hline
\multicolumn{6}{l}{$^*$ Calculated by authors}
\end{tabular}
\end{table}

\begin{figure}[ht]
\begin{center}
\includegraphics[width=86mm]{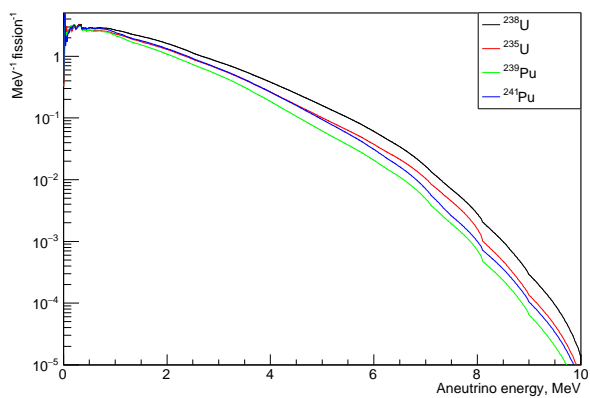}
\end{center}
\caption{\label{fig3} Antineutrino spectra calculated using modified fragments data base.}
\end{figure}

\begin{figure}[t]
\begin{center}
\includegraphics[width=86mm]{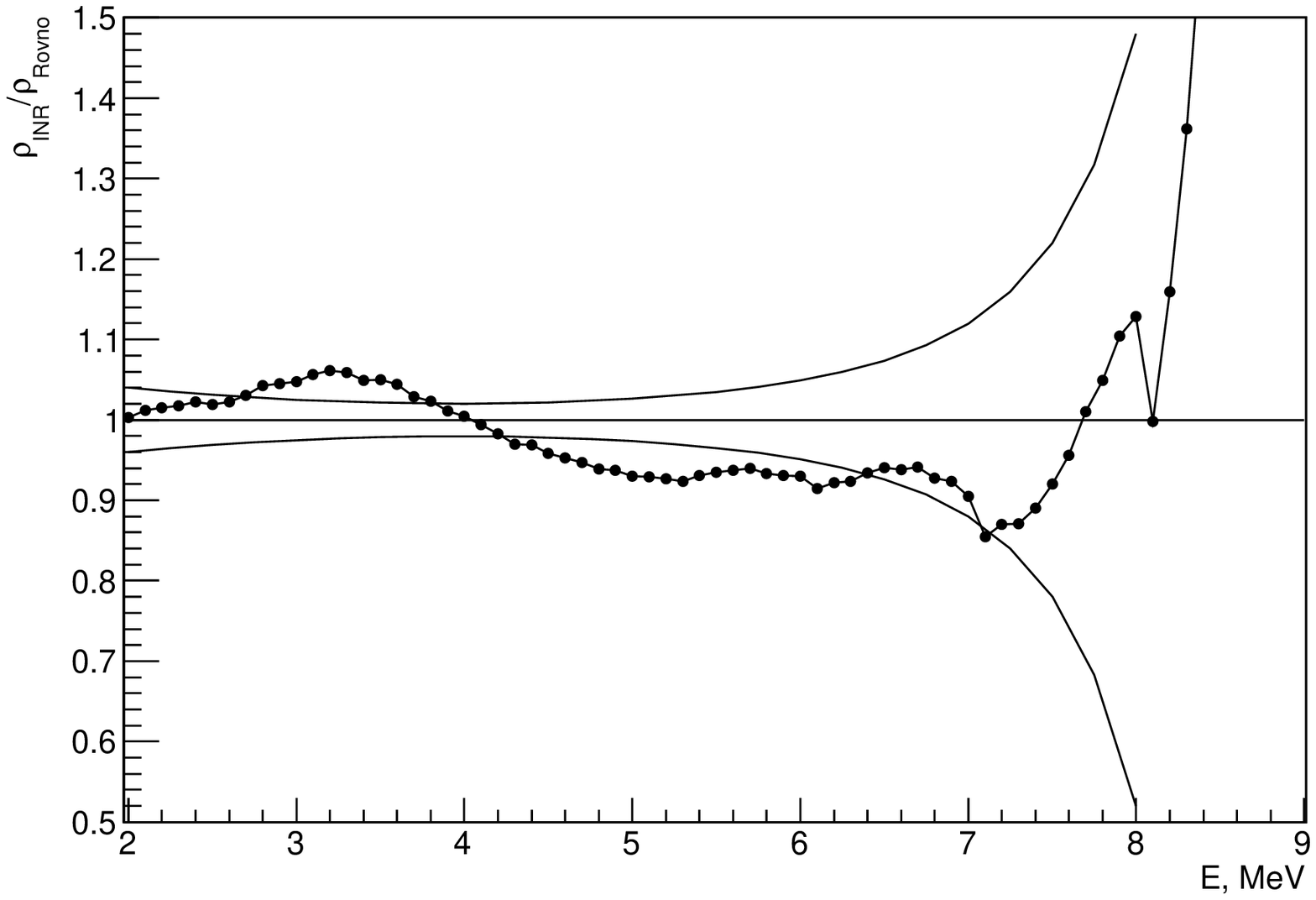}
\end{center}
\caption{\label{fig4} Ratio of INR calculated spectra mixture in proportion of Rovno experiment fuel composition to the spectrum of Rovno experiment according to the formula (5). One standard deviation corridor for experimental spectrum is shown.}
\end{figure}

Figure~\ref{fig3} shows the calculated antineutrino spectra with a modified base of fission fragments for $^{235}$U, $^{238}$U, $^{239}$Pu and $^{241}$Pu. We compared the spectrum mixed from calculated ones with the composition of the reactor core of the Rovno experiment with the experimental antineutrino spectrum in the form of formula (5) from \cite{klimov}. The ratio of the mixture of our spectra for a given core composition to spectrum (5) is shown in Figure~\ref{fig4}. The spectrum in the form of a formula is a smooth function and describes the behavior of the reactor spectrum on average, while the spectrum reconstructed from bins and the calculated one have a structure corresponding to the real spectrum.

\begin{figure}[ht]
\centering
\begin{minipage}{0.44\linewidth}
\includegraphics[width=\linewidth]{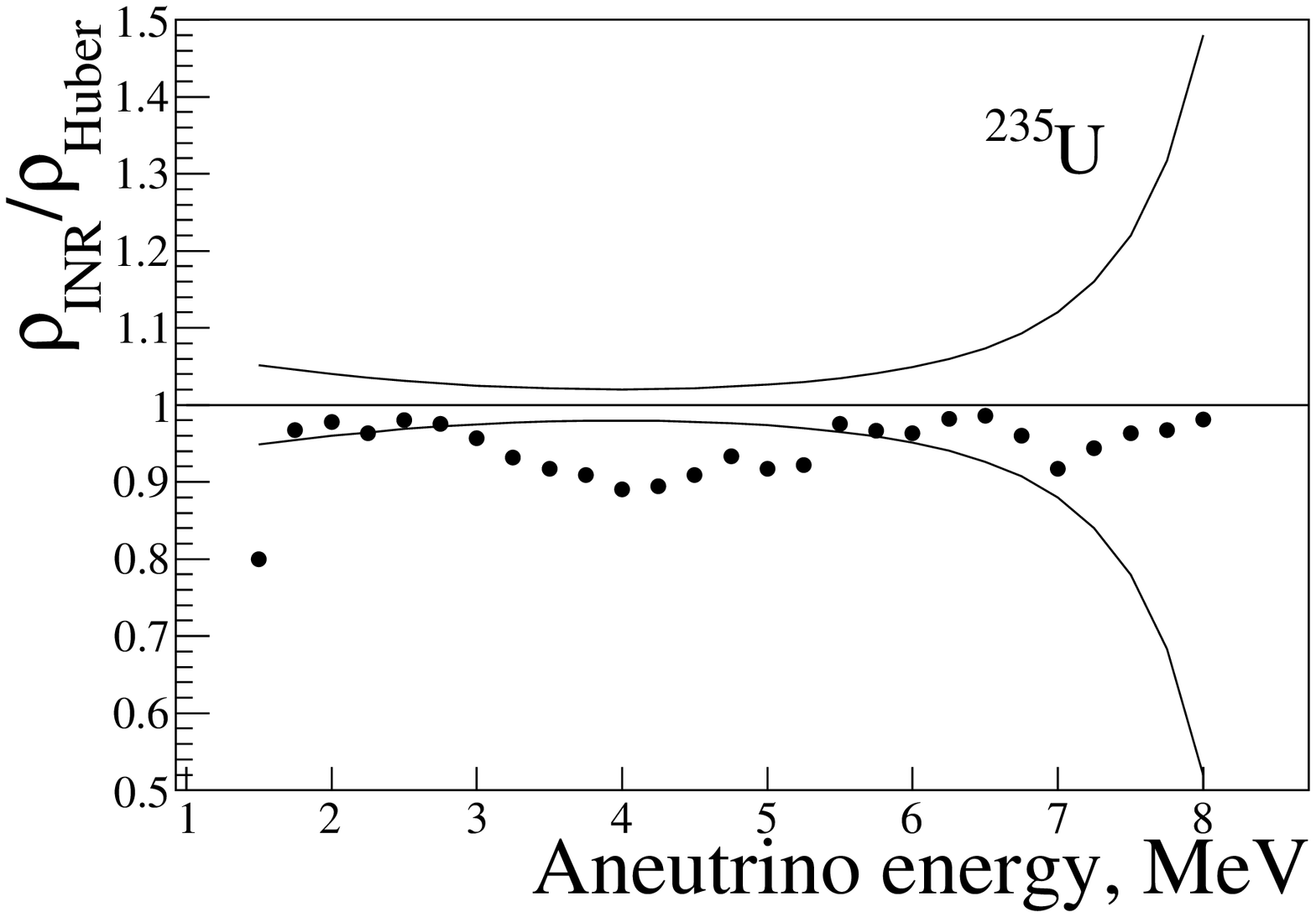}
\end{minipage}
\begin{minipage}{0.44\linewidth}
\includegraphics[width=\linewidth]{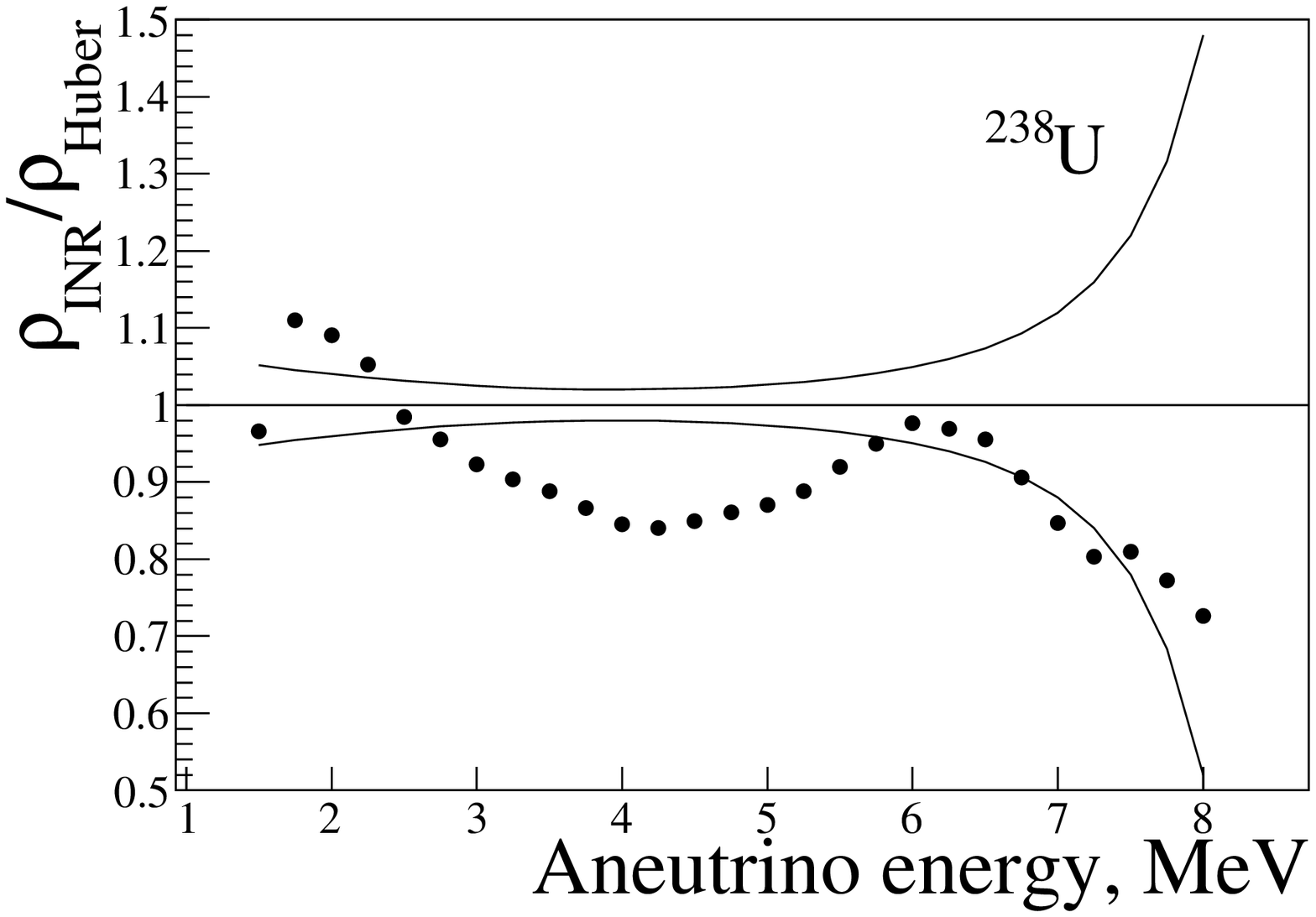}
\end{minipage}
\begin{minipage}{0.44\linewidth}
\includegraphics[width=\linewidth]{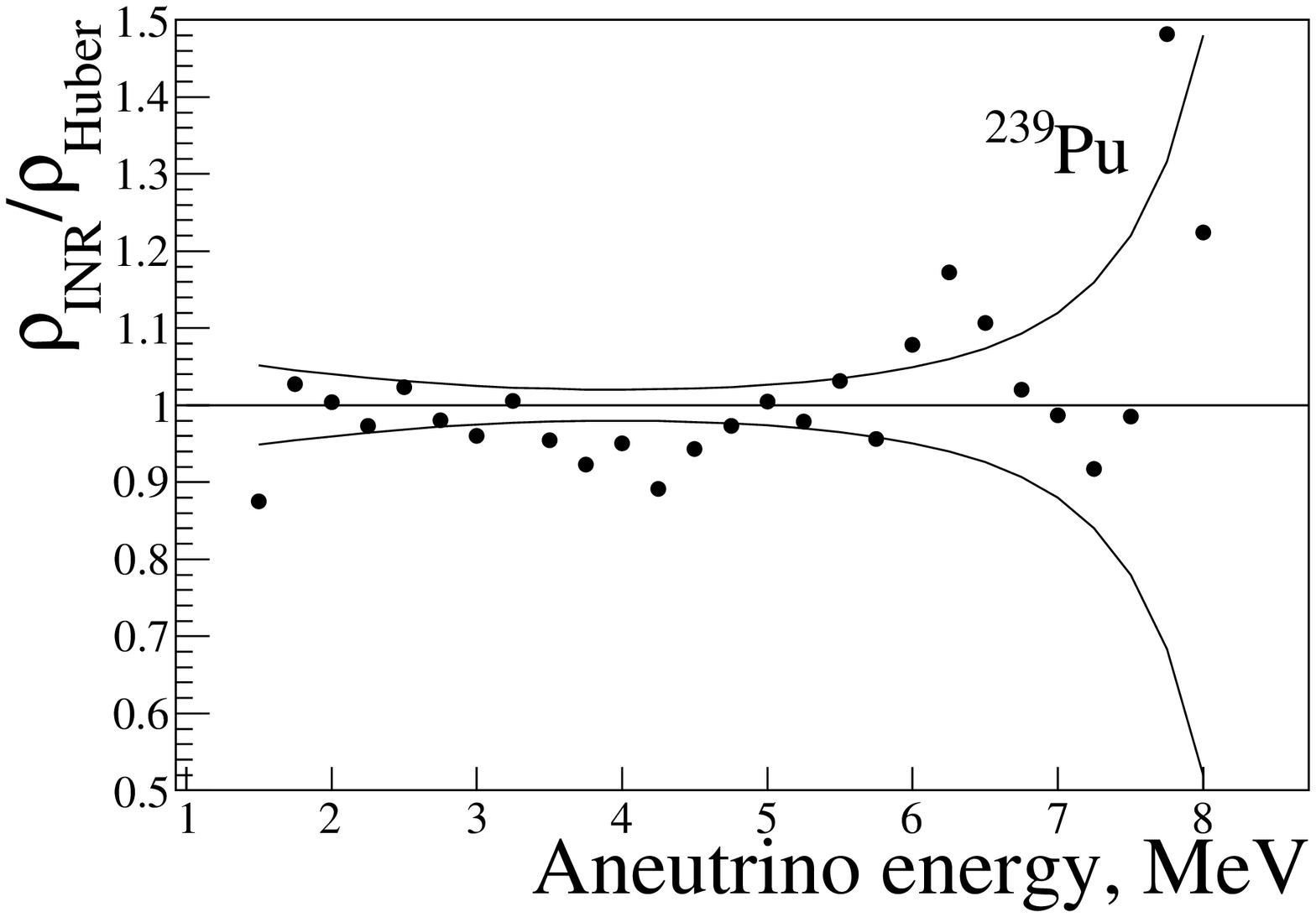}
\end{minipage}
\begin{minipage}{0.44\linewidth}
\includegraphics[width=\linewidth]{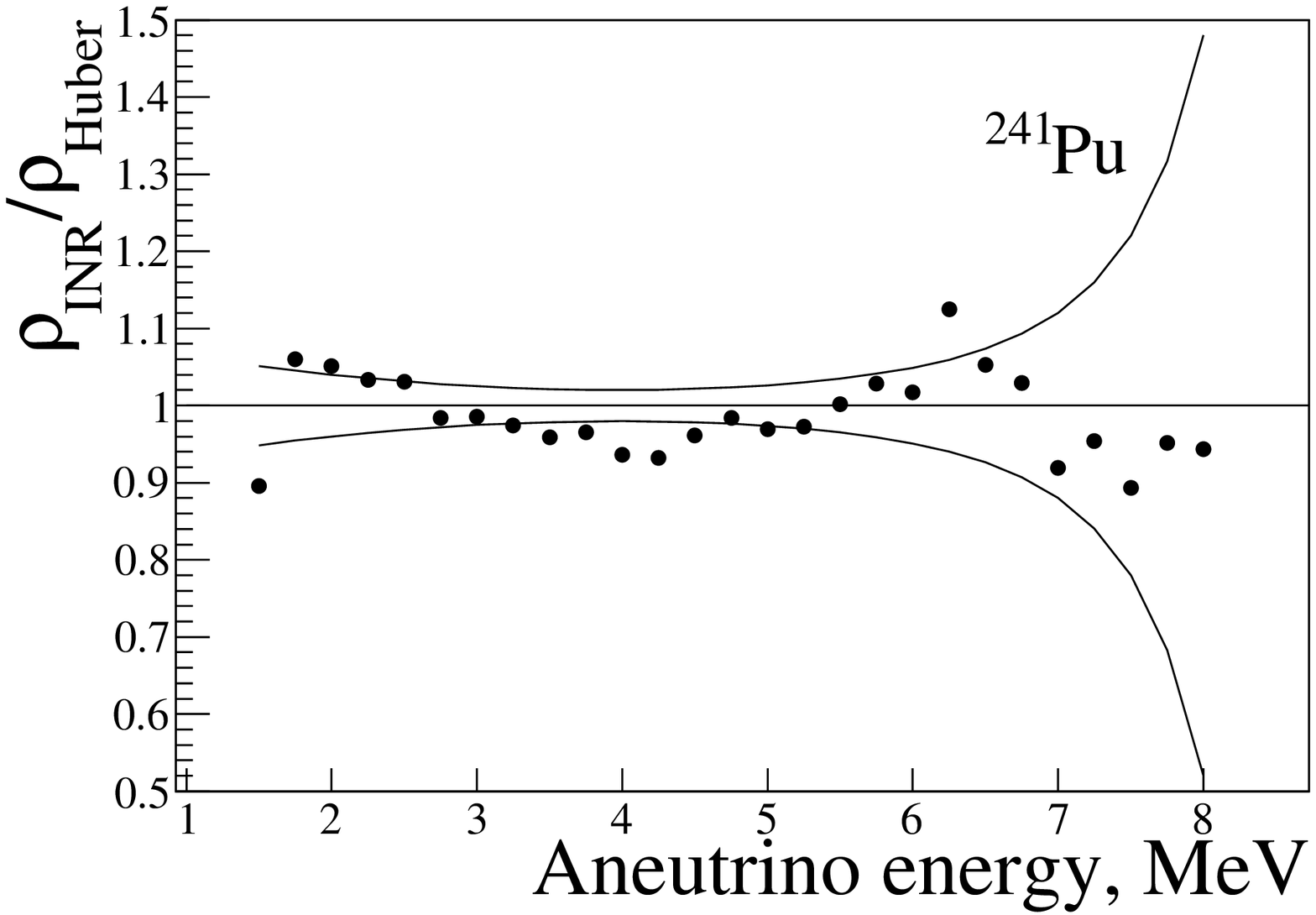}
\end{minipage}
\caption{\label{fig5} Ratio of INR antineutrino spectra to the ones from \cite{muell} and \cite{huber}.}
\end{figure}

Figure~\ref{fig5} shows the ratio of the antineutrino spectra of individual isotopes to the currently popular Huber-Muller spectra from \cite{muell}, \cite{huber}.

\begin{figure}[t]
\begin{center}
\includegraphics[width=86mm]{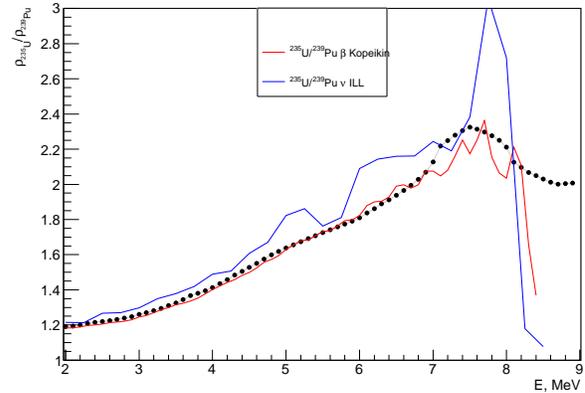}
\end{center}
\caption{\label{fig6} Ratio of $^{235}$U and $^{239}$Pu antineutrino and beta spectra shown. Black points $-$ INR spectra (this work), red line $-$ $\beta$-spectra ratio from the experiment at NRC Kurchatov Institute \cite{titov}, blue line $-$ $\bar{{\nu}_e}$-spectra ILL \cite{schre},\cite{hahn}.}
\end{figure}

Figure~\ref{fig6} shows the ratio of the spectra of antineutrinos from $^{235}$U and $^{239}$Pu. The ratios of beta spectra for the same isotopes from work by Kopeikin et al. \cite{titov} and antineutrino spectra from ILL \cite{schre}, \cite{hahn} are also shown here. At most energies, the ratio of our spectra coincides with the ratio of \cite{titov}. The observed difference in the energy range above 8 MeV can be explained by the edge effect, which takes place in the experiment for high-energy electrons. For example, in the ILL experiment, all three spectra after 8.5 MeV look very similar. This is possible because of the finite diameter of the tube, along which the electrons were pulled by a magnetic field from reactor core to the spectrometer, the more energetic ones were scattered on the walls and lost energy.
The ratio of the cross sections calculated from our spectra $\sigma_{f}$($^{235}$U)/ $\sigma_{f}$($^{239}$Pu = 1.45 appeares the same as in \cite{titov}.

\section{Discussion}

The calculated cross sections for the IBD reaction using antineutrino spectra from a number of works are given in Table \ref{tabl:sections}. The last column of this Table gives the averaged cross section for the IBD reaction corresponding to the core composition of Double Chooz experiment \cite{kerret}. Double Chooz \cite{kerret} obtained the experimental value of the IBD reaction cross section with an accuracy of 1\% for the core composition: $^{235}$U $–$ 0.52, $^{238}$U $–$ 0.087, $^{239}$Pu $–$ 0.333, and $^{241}$Pu $–$ 0.06. $^{DC}\sigma_{f} = (5.71 \pm 0.06) \times 10^{-43}$ cm$^2$/fission. In the experiment at Rovno \cite{klimov}, the measured cross section uncertainty was much worth $^{Rov}\sigma_{f} = (6.0 \pm 0.3)\times 10^{-43}$ cm$^2$/fission for the core composition: $^{235}$U $–$ 0.586, $^{238}$U $–$ 0.075, $^{239}$Pu $–$ 0.292 and $^{241}$Pu $–$ 0.047, but the measured cross sections are in good agreement within the experimental error and if corrected for the core composition. Until 2020, the most accurate measurement of the cross section for the IBD reaction was in the Bugey-4 experiment \cite{declais}. For a long time, the Bugey-4 experiment was the benchmark for other experiments, with a measurement error of 1.4\% $^{Buge}\sigma_{f} = (5.752 \pm 0.081) \times 10^{-43}$ cm$^2$/fission for the core composition: $^{235}$U $–$ 0.538, $^{238}$U $–$ 0.078, $^{239}$Pu $–$ 0.328 and $^{241}$Pu $–$ 0.056.

It can be seen from Table \ref{tabl:sections} that the Double Chooz cross section perfectly described by our calculated spectra and spectra converted from beta spectra measurements of ILL \cite{schre}, \cite{hahn}. The other cross sections differ from the experimental one by more than one Double Chooz standard deviation.

Table \ref{tabl:section2} shows the experimental cross sections with the best accuracy. The value of the Daya Bay experiment \cite{balant}, which has an accuracy of 2\%, was added to the cross sections with an accuracies of 1\% and 1.4\%.

\begin{table*}[ht]
\caption{Cross section value and its prediction on base of INR spectra for the most accurate measurements $\sigma_{f}\times10^{-43}$, cm$^{2}$/fission.}
\begin{minipage}{\textwidth}
\label{tabl:section2}
\centering
\vspace{2mm}
\begin{tabular}{ c | c | c | c | c | c | c | c }
\hline
\hline
  Experiment & \multicolumn{4}{|c|}{fuel content} & $^{i}\sigma_{f}$ & $^{INR}\sigma_{f}$ & $R$ \\
\cline{2-5}
 & $^{235}$U & $^{238}$U & $^{239}$Pu & $^{241}$Pu & & &\\
\hline
DC \cite{kerret}       & 0.520 & 0.087 & 0.333 & 0.060 & 5.71 $\pm$ 0.06 & 5.815 & 0.982 \\
Bugey-4 \cite{declais} & 0.538 & 0.078 & 0.328 & 0.056 & 5.752 $\pm$ 0.081 & 5.800 & 0.992 \\
Daya Bay \cite{balant} & 0.561 & 0.076 & 0.307 & 0.056 & 5.91 $\pm$ 0.12 & 5.836 & 1.013 \\
\hline
\hline
\end{tabular}
\end{minipage} \hfill
\end{table*}

Recently, an experiment at research reactor of the National Research Center "Kurchatov Institute" was performed having a goal to measure the ratio of the spectra of $^{235}$U and $^{239}$Pu \cite{titov}. Based on these measurements, the spectra of $^{235}$U and $^{238}$U were estimated. The result of calculations of cross sections for these spectra is given in the last row of Table \ref{tabl:sections}. The result of section calculations performed by the authors in \cite{titov} is given with an asterisk. The ratio of our $^{235}$U and $^{239}$Pu antineutrino spectra coincides with the ratio of the beta spectra of the same isotopes obtained in \cite{titov}.

$^{238}$U spectrum for a long time was used in estimations as calculated one. In 2013 this spectrum was obtained for the first time by using the same method as in \cite{schre}, \cite{hahn}. The beta spectrum of $^{238}$U was measured by using the fast neutron flux, and then it was converted into the antineutrino one. The obtained spectrum agrees with the spectrum of work \cite{muell}. The cross section turned out to be close to the Vogel spectrum data \cite{vogel} and one from this work.

\section{Conclusion}

We have presented a new calculation of the antineutrino spectra of fissile isotopes of uranium and plutonium, based on our upgraded database of fission fragments. In the available database for fragments with unknown decay schemes, the strength function was used to describe the probabilities of nuclear beta transitions. The strength function was chosen for the best description of the antineutrino spectra obtained at Rovno experiment.

The question of the need to use the Fermi function in the calculation of individual antineutrino spectra of fission fragments was investigated. Experimental antineutrino spectra are better described without using the Fermi function. If the Fermi function is not used to calculate the antineutrino spectrum, the individual spectra of antineutrino and electron turn out to be not exactly mirrored. The question of the need to check the symmetry of beta and antineutrino spectra was raised in \cite{silaeva}. This question remains open and requires a deeper analysis.

The antineutrino spectra obtained in our work are compared with the spectra \cite{muell}, \cite{huber} and experimental spectra \cite{sinev}. The spectra \cite{muell}, \cite{huber} show a difference in the region of 6 MeV of the neutrino energy, which is characteristic for the experimental spectra in the region of 5 MeV of the observed energy. Good agreement with the spectra from \cite{sinev} is observed, except for the soft region, where the insufficient removal of the response function and the influence of the spent fuel spectra could have an effect.

Comparison of the ratio of $^{235}$U and $^{239}$Pu antineutrino spectra agrees well with the measurement of beta particles spectra ratio for the same isotopes performed at the Kurchatov Institute. The calculated cross sections ratio for $^{235}$U and $^{239}$Pu spectra give the same ratio as in the work \cite{titov} of the National Research Center "Kurchatov Institute". The value of the cross sections ratio for uranium and plutonium $\sigma_{f}$($^{235}$U)/ $\sigma_{f}$($^{239}$Pu = 1.45.

The obtained strength functions for beta transitions of short-lived nuclei can be verified and refined by methods of nuclear physics.

\section*{Aknowledgments}

Authors are grateful to L.B. Bezrukov for useful discussions and valuable advices.

\end{document}